\documentclass[aps,twocolumn,superscriptaddress,amsmath,amssymb,amsfonts,longbibliography,floatfix]{revtex4-1}

\usepackage{soul}

\usepackage[utf8]{inputenc}
\usepackage{graphicx} 

\usepackage[colorlinks=true,citecolor=blue,linkcolor=blue,urlcolor=blue]{hyperref}

\usepackage[T1]{polski}
\usepackage[english]{babel}

\usepackage{bbold}

\usepackage{accents}
\newcommand\thickbar[1]{\accentset{\rule{.4em}{.8pt}}{#1}}

\graphicspath{{figures/}{./figures/}{.}}

\begin{document}

\title{
Influence of long-range interaction on Majorana zero modes
}

\author{Andrzej Wi\k{e}ckowski}
\email[e-mail: ]{andrzej.wieckowski@pwr.edu.pl}
\affiliation{
Department of Theoretical Physics, 
Faculty of Fundamental Problems of Technology,
Wroc\l{}aw University of Science and Technology,
PL-50370 Wroc\l{}aw, Poland}

\author{Andrzej Ptok}
\email[e-mail: ]{aptok@mmj.pl}
\affiliation{
Institute of Nuclear Physics, Polish Academy of Sciences, 
ul. W. E. Radzikowskiego 152, PL-31342 Krak\'{o}w, Poland}

\begin{abstract}
Majorana bound states, with their non-Abelian properties, are candidates for the realization of fault-tolerant quantum computation.
Here we study the influence of long-range many-body interactions on Majorana zero modes present in Kitaev chains.
We show that these interactions can suppress the lifetime of the Majorana zero mode.
We also discuss the role of long-range interactions on the Majorana state's spatial structure, and the overlap of the Majorana states localized at opposite ends of the chain.
We have determined that increasing the interaction strength leads to decreasing of the stability of the Majorana modes.
Moreover, we found out that interaction between particles located at more distant sites plays a more destructive role than the interaction between nearest neighbors.
\end{abstract}

\maketitle

\section{Introduction}

Kitaev's famous proposal~\cite{kitaev.01} for the realization of Majorana quasiparticles opened a period of intense study into these topological bound states~\cite{aguado.17,lutchyn.bakkers.18,pawlak.hoffman.19}.
Currently, realizations of Majorana bound states are expected in a few low-dimensional platforms.
First are semiconducting-superconducting hybrid nanostructures~\cite{deng.yu.12,mourik.zuo.12,das.ronen.12,finck.vanharlingen.13,
nichele.drachmann.17,gul.zhang.18,deng.vaitiekenas.16,deng.vaitiekenas.18}, where the interplay between intrinsic spin--orbit coupling, induced superconductivity, and external magnetic field lead to the realization of zero-energy bound states~\cite{lutchyn.bakkers.18}.
Second are one-dimensional (1D) chains of magnetic atoms deposited on a superconducting surface~\cite{nadjperge.drozdov.14,pawlak.kisiel.16,feldman.randeria.16,ruby.heinrich.17,jeon.xie.17,kim.palaciomorales.18}, where Majorana bound states are expected to form as a consequence of the magnetic moments in mono-atomic chains~\cite{klinovaja.stano.13,braunecker.simon.15,kaladzhyan.simon.17,andolina.simon.17}.
More recently, a realization of these zero-energy bound states in two-dimensional topological superconducting domains~\cite{drozdov.alexandradinata.14,menard.guissart.17,palaciomorales.mascot.18} and nanostructures with spin textures~\cite{yang.stano.16,mascot.cocklin.18,garnier.mesaros.19} have also been illustrated as well.

An essential property of Majorana zero modes (MZMs) are their non-Abelian statistics~\cite{nayak.simon.08}.
Such peculiarity makes them a very promising platform for the realization of fault-tolerant quantum computing~\cite{akhmerov.10,liu.wong.14,sarma.freedman.15,aasen.hell.16,hoffman.schrade.16,alicea.oreg.11}.
Quantum computations can be realized using {\it braiding protocols}~\cite{alicea.oreg.11,kim.tewari.15,beenakker.19,wieckowski.mierzejewski.19,trif.simon.19}, which can be practically implemented in wire-type systems~\cite{fatin.matosabiague.16}.
Here quantum qubit registers are stored in spatially separated MZMs, which are topologically protected from noise and decoherence~\cite{ivanov.01,cheng.lutchyn.12}.
The localized Majorana modes can also be manipulated by solely acting on the quantum dots~\cite{ruiztijerina.vernek.15,deng.vaitiekenas.16,ptok.kobialka.17,liu.sau.17,
prada.aguado.17,chevallier.szumniak.18,deng.vaitiekenas.18,reeg.dmytruk.18}.
For practical applications, it is crucial to describe the source of the decoherence in the system.
This is due to the fact that any decoherence can lead to additional errors in the state's coding~\cite{li.coish.18}.
From this, we should maximally clear out the source of decoherence in the system~\cite{nag.sau.18,zhang.mei.19}, which can be induced, e.g., by fluctuations~\cite{schmidt.rainis.12,lai.yang.18}.

In the context of the practical implementation of quantum computers based on MZMs, the interaction introduced in the system plays an important role in the computation process.
The stabilization of MZM can be achieved by introducing limited interaction strengths~\cite{stoudenmire.alicea.11,hassler.schuricht.12,gergs.niklas.16,dominguez.cayao.17}.
On-site repulsive interactions, in the half-spin fermion chain, was earlier discussed within in the Hartree--Fock approximation~\cite{stoudenmire.alicea.11,manolescu.marinescu.14}.
This type of interaction can lead to the decreasing of the Zeeman energy minimum value needed for MZM emergence.
Additionally, on-site interactions can stabilize the MZM~\cite{peng.pientka.15,dominguez.cayao.17}.
Long-range interactions, however, can reduce the decoherence rate~\cite{ng.15}. 
In the context of spinless fermions, interactions between nearest sites have been discussed using density-matrix renormalization-group (DMRG) methods~\cite{thomale.rachel.13,gergs.niklas.16}.
Moreover, in this case, moderate repulsive interactions stabilize the topological order.
In the present work, we study the influence of long-range interactions on the MZM's lifetime and spatial structures using exact diagonalization (ED) for the Kitaev chain.
The paper is organized as follows:
In Sec.~\ref{sec.model}, we introduce the microscopic model and present computational details.
In Sec.~\ref{sec.num_res}, we describe the numerical results.
Finally, we summarize the results in Sec.~\ref{sec.sum}.


\section{Model and methods}
\label{sec.model}

We consider a spinless fermion chain with $L$ sites, described by the Kitaev model~\cite{kitaev.01} extended by many-body interactions.
The system can be represented by the following Hamiltonian:
\begin{eqnarray}
\label{eq.ham}
\nonumber \mathcal{H} &=& \sum_{i=1}^{L-1} \left( - t a_i^{\dagger} a_{i+1}^{\phantom{\dagger}} + \Delta a_i^{\dagger} a_{i+1}^{\dagger} + \textrm{h.c.} \right) \\
&& - \mu\sum_{i=1}^L \widetilde n_i + \sum_{r=1}^{L-1}V_r \sum_{i=1}^{L-r} \widetilde n_i \widetilde n_{i+r} ,
\end{eqnarray}
where $a_i^\dagger$ ($a_i^{\phantom{\dagger}}$) is fermionic creation (annihilation) operator of spinless fermion at site $i$, while $\widetilde n_i = a_i^\dagger a_i^{\phantom{\dagger}}-1/2$.
Here $t$ is the hopping integral, $\Delta$ is the superconducting gap, $\mu$ is the chemical potential, and $V_r$ is the $r$-nearest neighbor interaction strength.

In the absence of the interactions, when $|\Delta|>0$, in the Kitaev model, 
there are two distinguished phases: topological and trivial~\cite{kitaev.01}.
Then in the thermodynamic limit it can be shown that the topological phase is present for $|\mu|\le 2t$ and the trivial phase is present for $|\mu|> 2t$.
MZMs can emerge only in the topological phase.
Here, one should notice that for a system with many-body interactions, the expression for the phase boundary can be more complicated~\cite{wieckowski.maska.18,thomale.rachel.13,katsura.schuricht.15}.
There are several methods for studying the presence of MZM in the system with many-body interactions.
Additionally, there exist a few indicators, which can be used for checking if the system is in the topological phase~\cite{gergs.niklas.16}.

From a theoretical point of view, studying quantum systems with many-body interactions is a relatively difficult task.
For example, DMRG methods allow for studying systems with thousands of sites, but are limited to short-range interactions. 
Here, we used ED for solving the chain with $L$ sites. 
Unfortunately, only small systems can be solved exactly (with $L\sim20$)~\cite{kozarzewski.mierzejewski.19}.
This method allows us to study all possible $r$ for selected system size.
For simplification and without loss of generality we take $r$ up to 4, due to the fact that for ED-based methods, only small system sizes are available.

MZMs, as states which are indistinguishable from their \textit{anti}states, should fulfill a few conditions.
Each MZM is equivalent to a zero mode $\Gamma$, which is fermionic operator satisfying the following relations~\cite{sarma.freedman.15}:
\begin{equation}
\Gamma^2 = 1,\qquad [\Gamma,\,\mathcal H] = 0 . \label{eq.majoranadef}
\end{equation}
In actual physical systems with finite length, in which one can realize a MZM, the second condition of~(\ref{eq.majoranadef}) is associated with the exponential suppression of energy splitting, i.e. $[\Gamma,\mathcal H] \propto e^{-L/\xi}$~\cite{sarma.freedman.15}, where $L$ is system size and $\xi$ is correlation length. 
MZM in the finite length system has nearly-zero energy and, in consequence, a finite lifetime~\cite{kells.15,wieckowski.maska.18}, which is observed experimentally~\cite{albrecht.higginbotham.16}.

To analyze the influence of long-range interactions into the spatial structure of the Majorana modes, we can analyze the $\Gamma$ modes in the Majorana basis representation.
Then MZMs in the Majorana operator $\gamma_i^+=a_i+a_i^\dagger,\gamma_i^-=i(a_i-a_i^\dagger)$ basis~\cite{kitaev.01} can be expressed as:
\begin{equation}
\Gamma =\sum_{i=1}^{2L} \alpha_i\gamma_i =\sum_{i=1}^L \left(\alpha_i^+ \gamma_i^+ + \alpha_i^- \gamma_i^-\right),
\end{equation}
where $\alpha^\pm_i$ are real coefficients.
We assume a normalization condition $\sum_i \alpha_i^2 = 1$ when $\Gamma^2=1$~\cite{wieckowski.maska.18}.
Here, $\gamma^{+}$ and $\gamma^{-}$ can be understood as new orthogonal basis matrices.
Moreover, this basis is a natural representation for the system which hosts MZM.

To check if MZMs can exist in our system, we used the same method which was introduced for studying integrals of motion in the Heisenberg model~\cite{mierzejewski.prosen.15,mierzejewski.prelovsek.15} and then was adapted for Kitaev model~\cite{wieckowski.maska.18}.
As the second condition for MZM is satisfied in the thermodynamic limit only (except fine-tuning parameters settings), we generate \textit{almost} conserved MZMs by solving the optimization problem~\cite{wieckowski.maska.18}:
\begin{equation}
\lambda = \max_{\{\alpha_i\}} \langle \thickbar \Gamma \thickbar \Gamma \rangle = \max_{\{\alpha_i\}} \sum_{ij}\alpha_i \langle \thickbar\gamma_i \thickbar\gamma_j\rangle \alpha_j,\label{eq.lambda}
\end{equation}
which becomes an eigenvalue problem for the matrix $\langle \thickbar\gamma_i \thickbar \gamma_j\rangle$.
We found the operator $\thickbar\Gamma$ averaged over time $\tau$ as \textit{close} as possible to the operator $\Gamma$. 
To measure the distance between operators $\thickbar \Gamma$ and $\Gamma$ we used the Hilbert--Schmidt operator norm defined in the following way: $\langle (\Gamma-\thickbar\Gamma)^2\rangle = \text{Tr}[(\Gamma-\thickbar\Gamma)^2]/\text{Tr}(\mathbb{1})$. 
We solved the eigenvalue problem for matrix $\langle \thickbar \gamma_i \thickbar \gamma_j\rangle$ and in result we obtained eigenvalues $\lambda$ with corresponding eigenvectors $[\alpha_i]$.
Averaging was done by high-oscillating terms cut-off in operator energy basis: 
\begin{equation}
\thickbar \Gamma = \sum_{nm} \theta\left( \frac1\tau - |E_n-E_m|\right)\langle n|\Gamma|m\rangle \,\,\, |n\rangle \langle m|,
\end{equation}
where $|n\rangle$ and $E_n$ are respectively eigenstate and eigenenergy of the Hamiltonian $\mathcal H$ and $\theta$ is the Heaviside function.
Such averaging, in the limit $\tau\to\infty$ is equivalent to calculating the following: $\lim_{\tau\to\infty} \frac1\tau \int_0^{\tau'}\text d\tau' \,\Gamma(\tau')$.
The value of $\lambda$ carries information about the distance between operators $\Gamma$ and $\thickbar \Gamma$.
In the limit $\tau\to\infty$, we can distinguish three different scenarios: $\lambda=1$, then $\Gamma$ is a strict integral of motion $\Gamma=\thickbar \Gamma$, the distance between them is zero and the MZM condition $[\mathcal H,\,\Gamma]=0$ is satisfied exactly. 
For $0<\lambda<1$, part of the information which is stored in $\Gamma$ is conserved. 
Finally, if $\lambda=0$, the information stored is lost.
In this work, we then look for the largest $\lambda$, and the corresponding eigenvectors $[\alpha_i]$, which carry information of the possible MZM realizations in the system.


\section{Numerical results}
\label{sec.num_res}

In this Section we present the numerical results. 
First, we start by describing the influence of the long-range interaction into Majorana modes lifetimes (Sec.~\ref{sec.lifetime}).
Next, we describe the spatial structure of the Majorana modes in the presence of the long-range interaction (Sec.~\ref{sec.profile}).

\begin{figure}[!t]
\centering
\includegraphics[width=\linewidth,keepaspectratio]{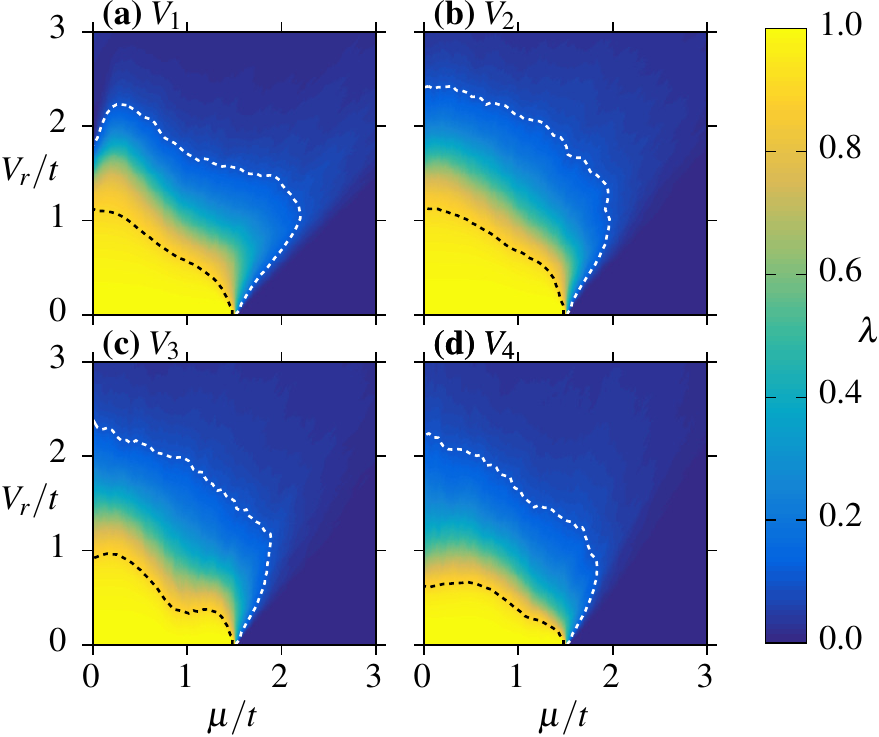}
\caption{
MZM correlation function $\lambda$ as a function of chemical potential $\mu$ and $r$-nearest neighbor interaction strength $V_r$ for
$\Delta/t=0.8$, $\tau=50$ and $L=10$.
Panels correspond to different long-range interaction $V_{r}$, as labeled.
Black and white contour marks $\lambda=0.9$ and 0.1, respectively.
}
\label{fig.plot1}
\end{figure}

\begin{figure}[!b]
\centering
\includegraphics[width=\linewidth,keepaspectratio]{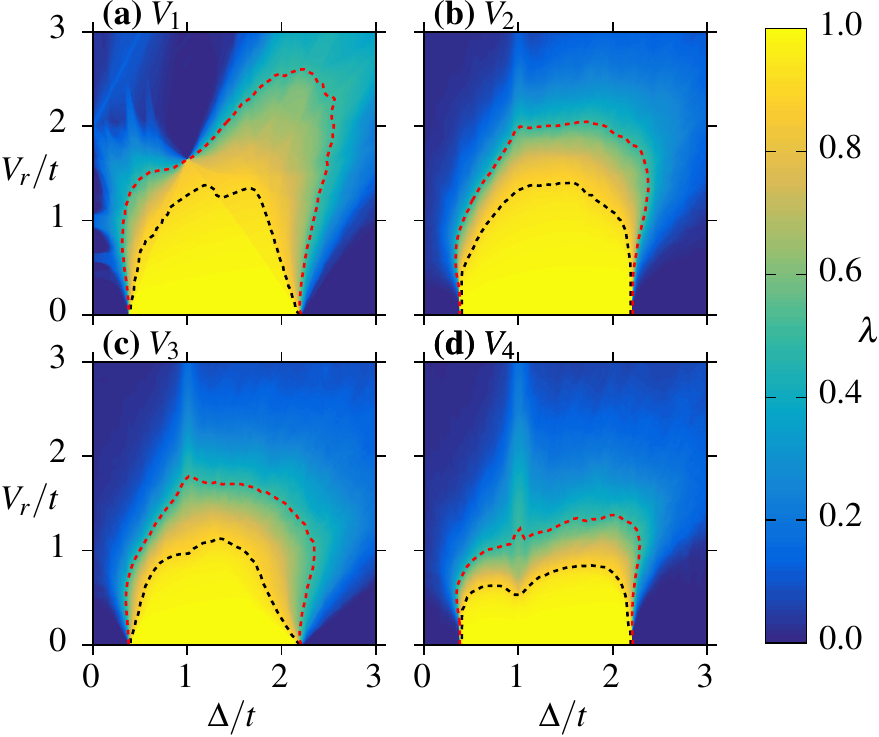}
\caption{
The same as in Fig.~\ref{fig.plot1}, but as a function of $\Delta$ and $V_r$. 
Black and red contour marks $\lambda=0.9$ and 0.5, respectively.
($\mu=0$)
\label{fig.plot2}
}
\end{figure}

\begin{figure}[!t]
\centering
\includegraphics[width=\linewidth,keepaspectratio]{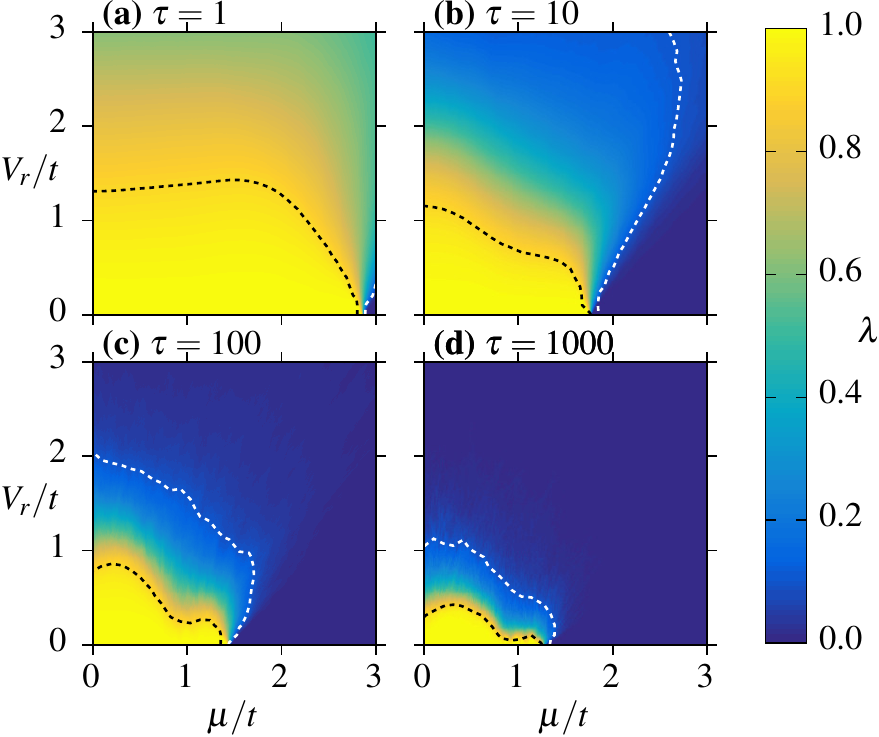}
\caption{
Finite time $\tau$ scaling.
The same as in Fig.~\ref{fig.plot1}(c), but for different times $\tau=1,10,100,1000$.
Black and white contour marks $\lambda=0.9$ and $0.1$, respectively.
\label{fig.plot3}
}
\end{figure}

\begin{figure}[!b]
\centering
\includegraphics[width=\linewidth,keepaspectratio]{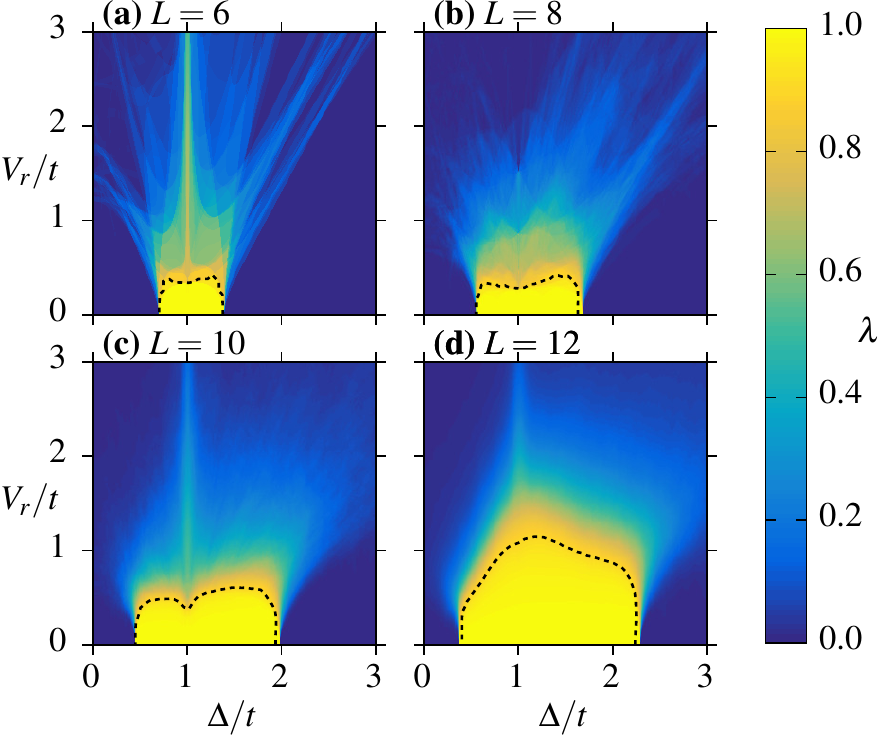}
\caption{ 
Finite size $L$ scaling.
The same as in Fig.~\ref{fig.plot2}(d), but for different system sizes $L=6,8,10,12$ and time $\tau=100$.
Black contour marks $\lambda=0.9$.
\label{fig.plot4}
}
\end{figure}

\subsection{Majorana modes lifetimes}
\label{sec.lifetime}

Results in Figs.~\ref{fig.plot1}--\ref{fig.plot4} show the most stable $\lambda$, which we found by solving Eq.~\eqref{eq.lambda}. Note that the biggest $\lambda$ is doubly degenerate -- for each MZM: $\Gamma^+$ and $\Gamma^-$, which are defined later.
To study the influence of long-range interaction, we study each $V_r$ independently, considering only one non-zero $r$-nearest neighbor interaction $V_r$ for given $r$.
In Fig.~\ref{fig.plot1} we compare influence of $V_r$ and $\mu$ on $\lambda$.
It is known that moderate interaction $V_1$ can lead to the broadening of the topological regime~\cite{stoudenmire.alicea.11}.
The same feature can be seen in our result, note the contour $\lambda=0.1$ in Fig.~\ref{fig.plot1}(a). 
However, this topological phase broadening is much smaller, when longer range interactions $V_2$, $V_3$ and $V_4$ are present in the system [Fig.~\ref{fig.plot1}(b)--Fig.~\ref{fig.plot1}(d)].
Moreover, increasing the interaction range $r$ decreases the area of strong MZM [see yellow area under the $\lambda=0.9$ contour in Fig~\ref{fig.plot1}(a)--Fig~\ref{fig.plot1}(d)].
Here we can notice that the transition from trivial to topological regime does not occur exactly at $|\mu|=2t$ in the case without interactions ($V_r=0$), due to a finite-size effect~\cite{kitaev.01}.

In Fig.~\ref{fig.plot2} we present the same as in Fig.~\ref{fig.plot1}, but as a function of $\Delta$, instead of $\mu$.
Again one can see the topological phase decreasing as the interaction range $r$ grows.
One can see here characteristic line along $\Delta/t=1$.
This fading line is related to the fact that Kitaev model for the non-interacting case $\Delta=|t|$ and $\mu=0$ (special parameter tweak) contains MZM, which are exactly integrals of motion even for finite system size~\cite{kitaev.01}.
It seems that for large $\tau$ and $\Delta/t\gg1$ MZM are absent in the system.
However, this is only a finite size effect, which we explained in detail in Fig.~\ref{fig.plot4}.

\begin{figure}[!t]
\centering
\includegraphics[width=\linewidth,keepaspectratio]{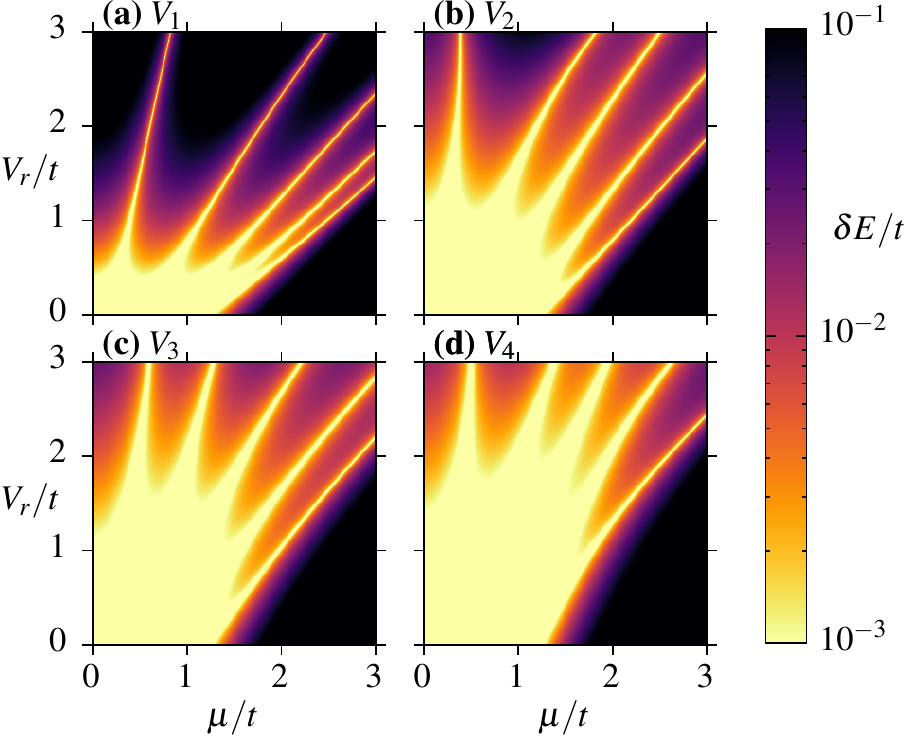}
\caption{
Ground-state degeneracy $\delta E$ as a function of interaction $V_r$ and chemical potential $\mu$.
Panels correspond to the different long-range interaction $V_{r}$, as labeled.
System parameters the same as in Fig.~\ref{fig.plot1}.
}\label{fig.plot5}
\end{figure}

\begin{figure}[!b]
\centering
\includegraphics[width=\linewidth,keepaspectratio]{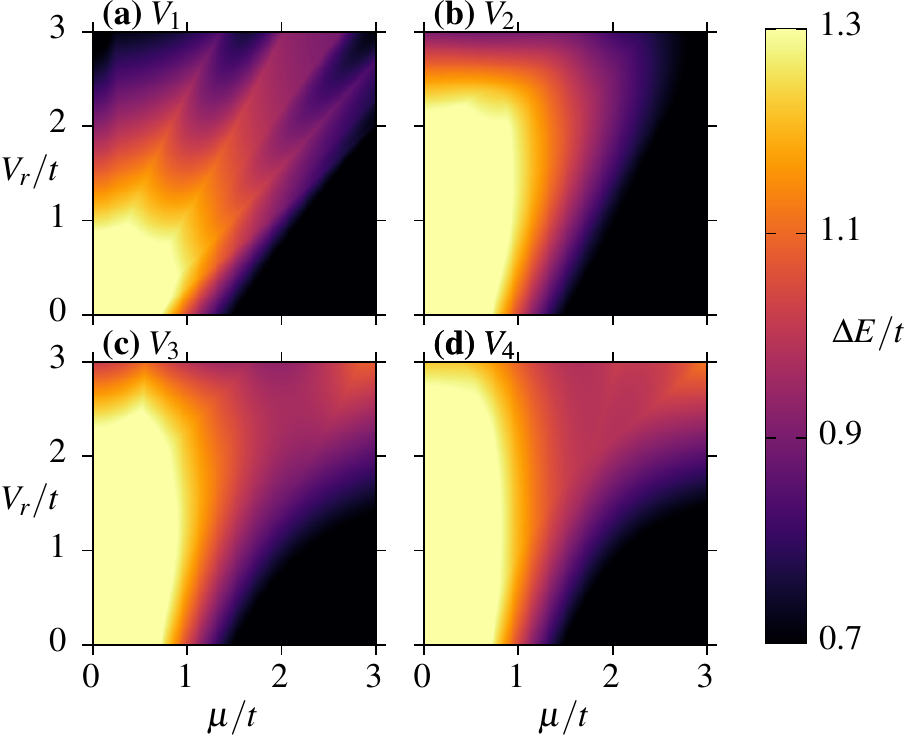}
\caption{
Energy gap $\Delta E$ as a function of interaction $V_r$ and chemical potential $\mu$.
Panels correspond to different long-range interaction $V_{r}$, as labeled.
System parameters same as in Fig.~\ref{fig.plot1}.
}
\label{fig.plot6}
\end{figure}

To study the topological phase in the thermodynamic limit $L\to\infty$ and $\tau\to\infty$ one should be extremely careful doing the size/time scaling.
Except for special parameter tweak, following limit always tends to zero: $\lim_{L\to\infty}\lim_{\tau\to\infty} \lambda = 0$.
In Fig.~\ref{fig.plot3} we show how $\lambda$ vanishes over time $\tau$ for the selected case.
However, there is a non-zero topological regime even for a large $\tau=1000$ and for a relatively small system with $L=10$.
In a contrast, limit $\Lambda=\lim_{L\to\infty}\lim_{\tau\to\infty} \lambda$ in general can be different than 0. 
The order of these limits is essential, for an almost strong MZM value of $\Lambda\simeq1$~\cite{wieckowski.maska.18}. 
In Fig.~\ref{fig.plot4}, a finite-size scaling is presented.
The procedure of extrapolation of $\Lambda$ can be found in Ref.~\cite{wieckowski.maska.18}.
However, in this work, to compare the influence of interaction range $r$, finite time results are sufficient for the discussion.

Next, we check the necessary condition for soft MZM, i.e. degeneracy of the ground-state energies $\delta E=\left|E_0^{\mathrm o}-E_0^{\mathrm e}\right|$ and spectral gap $\Delta E = \min\{E_1^{\mathrm e}-E_0^{\mathrm e},\,E_1^{\mathrm o}-E_0^{\mathrm o}\}$, where $E_n^{\mathrm e}$ ($E_n^{\mathrm o}$) is $n$-eigenenergy from even (odd) parity regime~\cite{ng.15}.
Ground-state degeneracy $\delta E$ and energy gap $\Delta E$ results for different interaction $V_r$ range $r$ can be found in Figs.~\ref{fig.plot5} and \ref{fig.plot6}, respectively.
Surprisingly, one may conclude from the results presented in Fig.~\ref{fig.plot5} that increasing the interaction range $r$ topological phase increases as a consequence of increasing the yellow regime, where $\delta E$ is small.
Simultaneously, in Fig.~\ref{fig.plot6}, the area with a bigger energy gap $\Delta E$ grows with the interaction range $r$.
It should be stressed that the $\delta E$ condition is necessary, but it is not sufficient.
In Fig.~\ref{fig.plot5}(a) one can identify a few yellow stripes.
These lines separate regions where the ground-state average particle number $\langle N \rangle = \langle \sum_i a_i^\dagger a_i^{\phantom{\dagger}}\rangle $ is close to an integer value: $0,1,\dots,L$~(see Supplementary Material for Ref.~\cite{wieckowski.maska.18}).
These lines are the consequence of energy level crossings and are not related to MZM presence in the system.

\subsection{Spatial structure of Majorana modes}
\label{sec.profile}

\begin{figure}[!t]
\centering
\includegraphics[width=\linewidth,keepaspectratio]{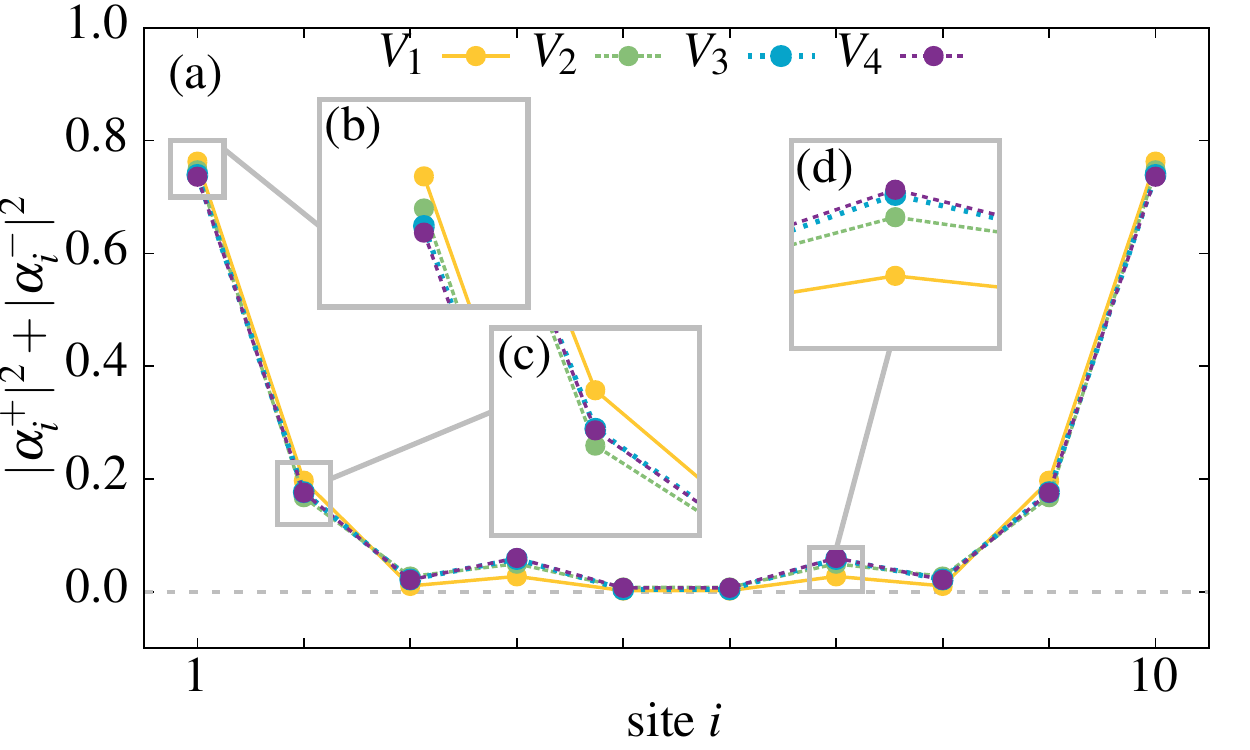}
\caption{
Spatial structure of the Majorana states $\Gamma^+$ and $\Gamma^-$. Results for $L=10$, $\Delta/t=0.4$, $V_r/t=1$, and $\mu/t=0.7$.
\label{fig.plot8a}
}
\end{figure}

To study the spatial profile of the MZM, we can express the $\Gamma$ state in the Majorana basis, which was included earlier (cf.~Sec.~\ref{sec.model}).
Then, we can find a pair of orthogonal operators $\Gamma^+$ and $\Gamma^-$:
\begin{equation}
\Gamma^\pm = \sum_{i=1}^L \alpha_i^\pm \gamma_i^\pm ,
\end{equation}
which describe a projection of the $\Gamma$ states into the pure--Majorana $\gamma$ states.
Because of this, every $\Gamma^\pm$ state contains only $\alpha_{i}^{\pm} \neq 0$ (describing contribution of the $\gamma^{\pm}$ state), while in the same time $\alpha_{i}^{\mp} =0$.
Examples of numerical results are shown in Fig.~\ref{fig.plot8a}, where $|\alpha_{i}^{+}|^{2} + |\alpha_{i}^{-}|^{2}$ is presented.
This quantity corresponds to the local density of states~\cite{matsui.sato.03} or differential conductance~\cite{chevallier.klinovaja.16}.
Additionally, in the case of a uniform chain, due to symmetry in the chain midpoint, the coefficients for $\Gamma^+$ and $\Gamma^-$ must be swapped in space, i.e., $\alpha_i^+=\alpha_{L+1-i}^-$. 
Using such constraint, one can generate coefficients only for one of $\Gamma^\pm$ to study the spatial structures.
As we can see, increasing interaction range $r$ leads to decrease of the MZM localization, i.e. when $r$ grows, the sum $|\alpha_{i}^{+}|^{2} + |\alpha_{i}^{-}|^{2}$ at the center of the chain increases [cf.~Fig.~\ref{fig.plot8a}(b)--Fig.~\ref{fig.plot8a}(d)].
At the same time, the value of this expression decreases at the ends of the chain.
Such behavior can be explained by decreasing the overlap between MZM located at both left and right ends of the chain.
In contrast, the interaction between nearest-neighbor sites leads to the stabilization increment of the MZM~\cite{dominguez.cayao.17,wieckowski.maska.18}.
Moreover, this emphasizes the importance of many-body interactions on the MZM lifetime.

\begin{figure}[!b]
\centering
\includegraphics[width=\linewidth,keepaspectratio]{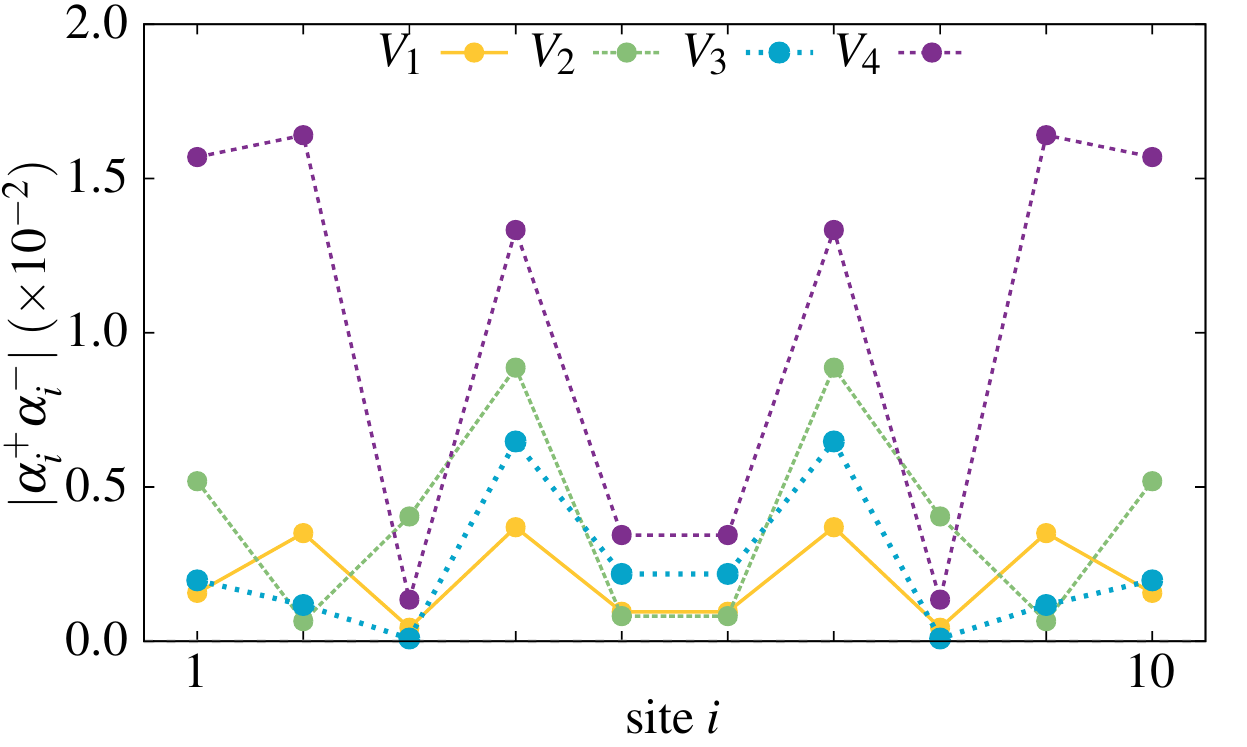}
\caption{
Local overlap $|\alpha_i^+\alpha_i^-|$ between MZM $\Gamma^+$ and $\Gamma^-$.
System parameters are the same as in Fig.~\ref{fig.plot8a}.
}
\label{fig.plot8b}
\end{figure}

\begin{figure}[!t]
\centering
\includegraphics[width=\linewidth,keepaspectratio]{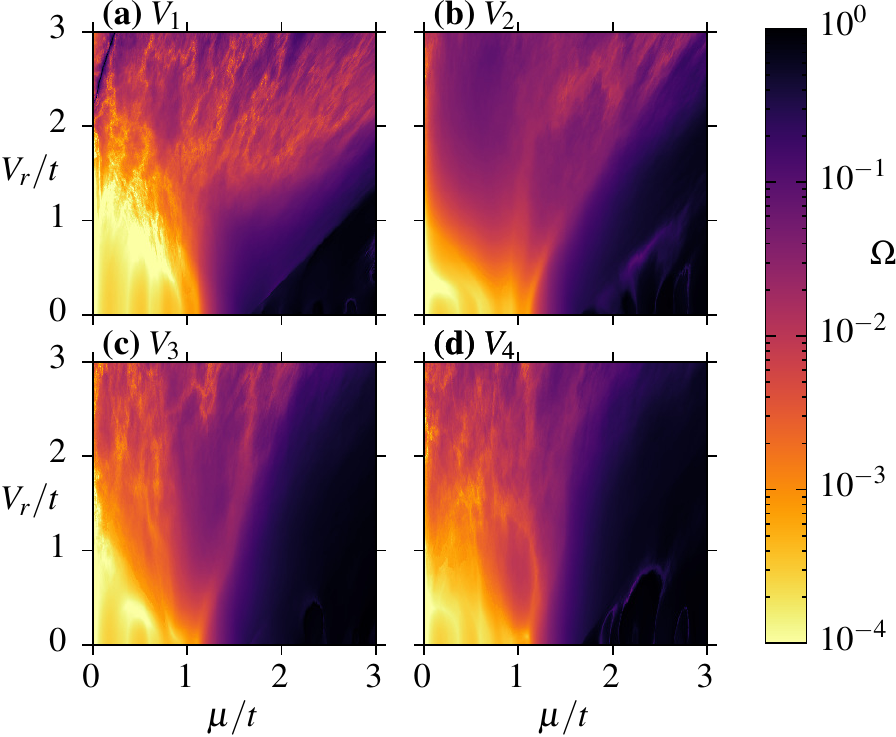}
\caption{
Overlap $\Omega$ between left, and right Majorana states, 
as a function of interaction $V_r$ and chemical potential $\mu$.
Panels correspond to different long-range interaction $V_{r}$, as labeled.
System parameters are the same as in Fig.~\ref{fig.plot1}.
}
\label{fig.plot01}
\end{figure}

\begin{figure}[!b]
\centering
\includegraphics[width=\linewidth,keepaspectratio]{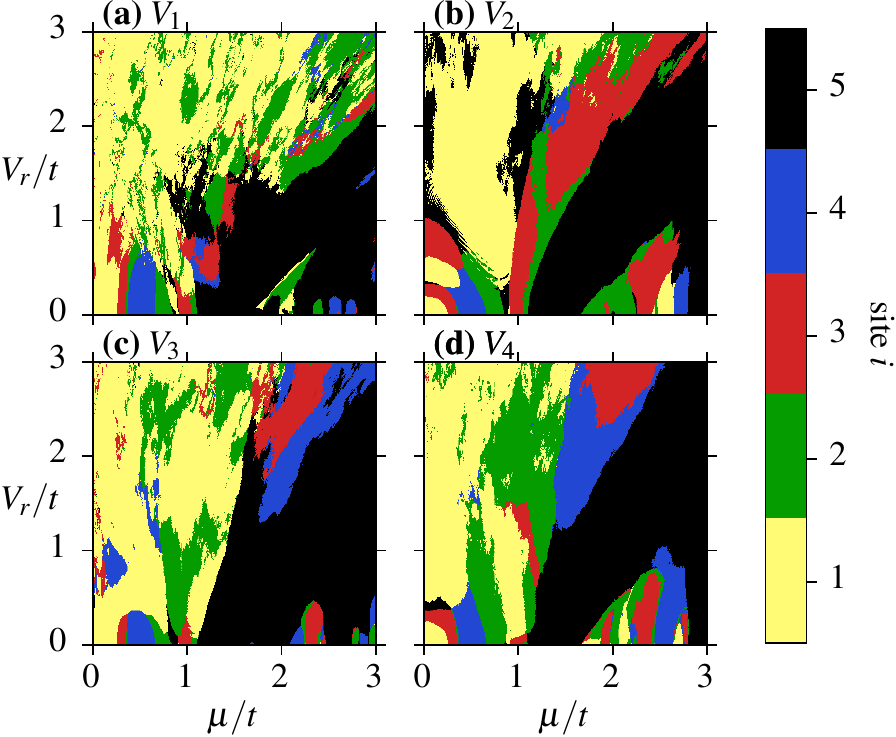}
\caption{
\label{fig.plot03}
The site number $i$ for which $|a_i^+ a_i^-|$ is maximal as a function of interaction $V_r$ and chemical potential $\mu$.
Panels correspond to different long-range interaction $V_{r}$, as labeled.
System parameters are the same as in Fig.~\ref{fig.plot1}.
}
\end{figure}

As a degree of non-locality of the two Majorana states, we can define their overlap~\cite{prada.aguado.17,deng.vaitiekenas.18}:
\begin{equation}
\Omega = \|\Gamma^+ \widetilde \Gamma^- \| = \sum_{i=1}^{L} | \alpha_{i}^{+} \alpha_{i}^{-} | ,
\end{equation}
where $\widetilde \Gamma^- = \mathcal U \Gamma^-\mathcal U^\dagger$ is reflection in space of $\Gamma^+$ and $\mathcal U$ is unitary operator described by transformation of basis matrices $\mathcal U \gamma_i^\pm\mathcal U^\dagger = \gamma_i^\mp$.
From the definition, $\Omega$ takes ranges from 0 (no overlap) to 1 (perfect overlap).
Here, we note that the $\Omega$ strongly depends on $L$~\cite{dumitrescu.roberts.15}.
Moreover, this quantity can be associated with the resilience of the Majorana qubit to local environmental noise, with complete non-locality $\Omega = 0$ denotes topological qubit protection~\cite{prada.aguado.17}.

In general, the overlap $\Omega$ can be controlled by some parameter modification, like electrostatic potential~\cite{ptok.cichy.18,penaranda.aguado.18,kobialka.ptok.19,rainis.trifunovic.13} or inter-site interactions~\cite{dominguez.cayao.17,wieckowski.maska.18}.
In our case, we control $\Omega$ by modification of the long-range interaction $V_r$ and the chemical potential $\mu$ in the whole system, for which the result is presented in Fig.~\ref{fig.plot01}.
For weak $V_{r}$ and doping $\mu$ overlapping is exponentially small. 
When interaction $V_r$ increases, $\Omega$ decreases -- this effect seems to be independent of interaction $V_r$ range.
As one can see, the MZM overlap $\Omega$ is more sensitive to controlling the chemical potential $\mu$ than interaction $V_r$ modifications.

Similar behaviour can be observed in Fig.~\ref{fig.plot03}, where we show site index $i$ for which the ``local'' overlapping $| a_{i}^{+} a_{i}^{-} |$ reaches maximal value.
As one can see, increasing chemical potential $\mu$ leads to a stronger overlap between Majorana states (at the center of the chain, i.e., $i=5$). 
In contrast, increasing of the long-range interaction $V_{r}$ leads to increasing overlapping near the edge of the chain -- the maximal value of overlap is more visible outside than in the center of the chain. 
Note that, fast changes of site index $i$ for $\mu/t \sim 1$ are associated with numerical accuracy, i.e. local overlap $|\alpha_i^+\alpha_i^-|$ is relatively small and comparable for all $i$ (small variance).


\section{Summary}
\label{sec.sum}

We have studied the influence of interaction range on the Majorana zero mode lifetime and spatial structure in the Kitaev chain.
The Majorana zero mode's lifetime is an important quantity from a practical point of view and can be related to the topological qubit decoherence time.
For the practical application of Majorana zero modes, one needs to extend the decoherence time, which will aid the realization of a quantum computers based on their non-Abelian properties.

From previous theoretical calculations based on DMRG methods, moderate repulsive interactions between the nearest sites can lead to the stabilization of the topological order~\cite{thomale.rachel.13,gergs.niklas.16}.
It should be emphasized that the dissipation and dephasing of the Majorana zero modes have also been studied in the presence of nearest neighbor interactions~\cite{ng.15}.
In this case, the dissipation and dephasing noises can induce parity- and non-parity preserving transitions.
Moreover, the dissipation and dephasing rates can be reduced by increasing the interaction strength at sufficiently low temperature, which can lead to extended coherence times for the Majorana mode~\cite{ng.15}.

In this paper, we have shown that long-range interaction strongly modifies the lifetime of the Majorana zero mode.
These interactions decrease the lifetime of the MZM.
Moreover, we have discovered that interaction between particles located at distant sites is more significant than the interaction between nearest neighbors.
This behavior can have a crucial role from the practical point of
view in real materials, where interaction decays with distance. 
This destructive character can be crucial for the practical implementation of Majorana zero modes as topological qubits.
This type of interaction leads to the overlap between two Majorana bound states localized at the opposite end of the chain.
Naturally, it can be a source of decoherence of these states.
In summary, to guarantee the efficiency of quantum computers based on Majorana zero modes, the suppression of the long-range interaction is required.

\begin{acknowledgments}
The authors are thankful to David J. Alspaugh, Szczepan G\l{}odzik, Jan \L{}a\.{z}ewskim, Marcin Mierzejewski, Pascal Simon, and Olga Sikora for very fruitful discussions and comments.
This work was supported by the National Science Centre (NCN, Poland) under Grant No. 2016/23/B/ST3/00647.
\end{acknowledgments}

\bibliography{biblio}

\end{document}